\documentclass{article}

    \usepackage[final]{neurips_2020}

\usepackage[utf8]{inputenc} 
\usepackage[T1]{fontenc}    
\usepackage{hyperref}       
\usepackage{url}            
\usepackage{booktabs}       
\usepackage{amsfonts}       
\usepackage{nicefrac}       
\usepackage{microtype}      
\usepackage[pdftex]{graphicx}

\usepackage{color}

\definecolor{grey}{rgb}{0.5, 0.5, 0.5}
\newcommand\spv[1]{\textcolor{grey}{#1}}

\title{Multi-Format Contrastive Learning of Audio Representations}

\author{%
  Luyu~Wang \\
  Google DeepMind\\
  \texttt{luyuwang@google.com} \\
  \And
  A\"aron van den Oord \\
  Google Deepmind \\
  \texttt{avdnoord@google.com} \\
}

\begin{document}

\maketitle

\begin{abstract}
  Recent advances suggest the advantage of multi-modal training in comparison with single-modal methods. In contrast to this view, in our work we find that similar gain can be obtained from training with different formats of a single modality. In particular, we investigate the use of the contrastive learning framework to learn audio representations by maximizing the agreement between the raw audio and its spectral representation. We find a significant gain using this multi-format strategy against the single-format counterparts. Moreover, on the downstream AudioSet and ESC-50 classification task, our audio-only approach achieves new state-of-the-art results with a mean average precision of 0.376 and an accuracy of $90.5\%$, respectively.
\end{abstract}

\section{Introduction}

Self-supervised learning leverages proxy tasks to learn useful representations of the data without requiring manually annotated labels. In computer vision, methods using contrastive losses \cite{hjelm2018learning, oord2018representation, bachman2019learning, henaff2020data, he2020momentum, chen2020simple, chen2020big} stand out on the ImageNet benchmark \cite{deng2009imagenet}, which learns by maximizing the similarity between augmented views from the same image. Contrastive learning also facilitates the recent rapid progress in unsupervised speech recognition \cite{schneider2019wav2vec, baevski2019vq, kawakami2020learning, riviere2020unsupervised, kahn2020libri, baevski2020wav2vec}. Good speech representations should be able to extract transient linguistic information. Therefore, these works rely on context prediction models that output representations with a fine-grain temporal resolution, while excluding non-speech sound that can distract the model from the task. For both image and speech recognition, the gaps between supervised and unsupervised representations have largely been eliminated.

Unlike speech recognition, the multi-instance audio events classification problem requires discriminative representations to tell the differences among a broad class of audio events. AudioSet \cite{gemmeke2017audio} is the ImageNet-scale dataset for general audio understanding, which contains 527 highly imbalanced event classes. Recent development in this direction has mainly been focusing on supervised learning \cite{hershey2017cnn, wang2019comparison, ford2019deep, kong2020panns}. 
In \cite{jansen2018unsupervised}, a triplet-based unsupervised approach is introduced to learn audio features from augmented spectrograms.
Later, the effectiveness of contrastive predictive coding (CPC) is invstigated in \cite{wang2020contrastive}, which operates on raw waveforms and is widely used for speech models.
Meanwhile, recent works show that better representations can be learned by the proxy task of predicting whether the visual and audio signals come from the same video \cite{arandjelovic2017look, arandjelovic2018objects, korbar2018cooperative, owens2018audio, jansen2019coincidence, alwassel2019self, alayrac2020self, morgado20avid, mandela2020datatrans}.
On the AudioSet benchmark, \cite{jansen2019coincidence} shows that it is beneficial to maximize the coincidence between video and audio with the contrastive loss. There is a clear advantage by further taking the additional text modality into account \cite{alayrac2020self}. However, the state-of-the-art unsupervised audio model is still lagging behind the supervised one \cite{kong2020panns} (mean average precision (mAP) 0.309 vs 0.439). 

Apart from the raw audio format, traditional signal processing allows us to convert the wavefroms into the spectral representations via shoft-time Fourier transforms (STFTs) \cite{jurafsky2008speech}. Such spectrograms can further be retrieved into log-mel filter banks and mel-frequency cepstral coefficients (MFCCs), which are the dominant formats for supervised and unsupervised audio recognition \cite{hershey2017cnn, wang2019comparison, ford2019deep, kong2020panns, jansen2018unsupervised, jansen2019coincidence}. However, to the best of our knowledge, all previous works consider only one format of the audio modality. In this paper, we investigate the use of contrastive learning to learn audio representations from multiple formats. Different from the models that drive the current progress on learning unsupervised speech representations \cite{schneider2019wav2vec, baevski2019vq, kawakami2020learning, riviere2020unsupervised, kahn2020libri, baevski2020wav2vec}, our method does not rely on the context prediction network, and directly contrast two augmented views (which resembles image models in \cite{he2020momentum, chen2020simple}). We conduct experiments on different input formats, architectures, and augmentations. It is found that much better representations can be learned by maximizing the agreement between two views from the same audio represented by the raw waveform and log-mel filterbanks. As a result, our single-modal model has a test mAP of 0.376 on AudioSet, outperforming the previous best multi-modal score of 0.309 \cite{alayrac2020self} by a large margin. Moreover, it generalizes to the ESC-50 downstream classification task with a new state-of-the-art accuracy of $90.5\%$.

\section{Learning framework}

\begin{figure}[t]
  \centering
  \includegraphics[width=0.4\linewidth]{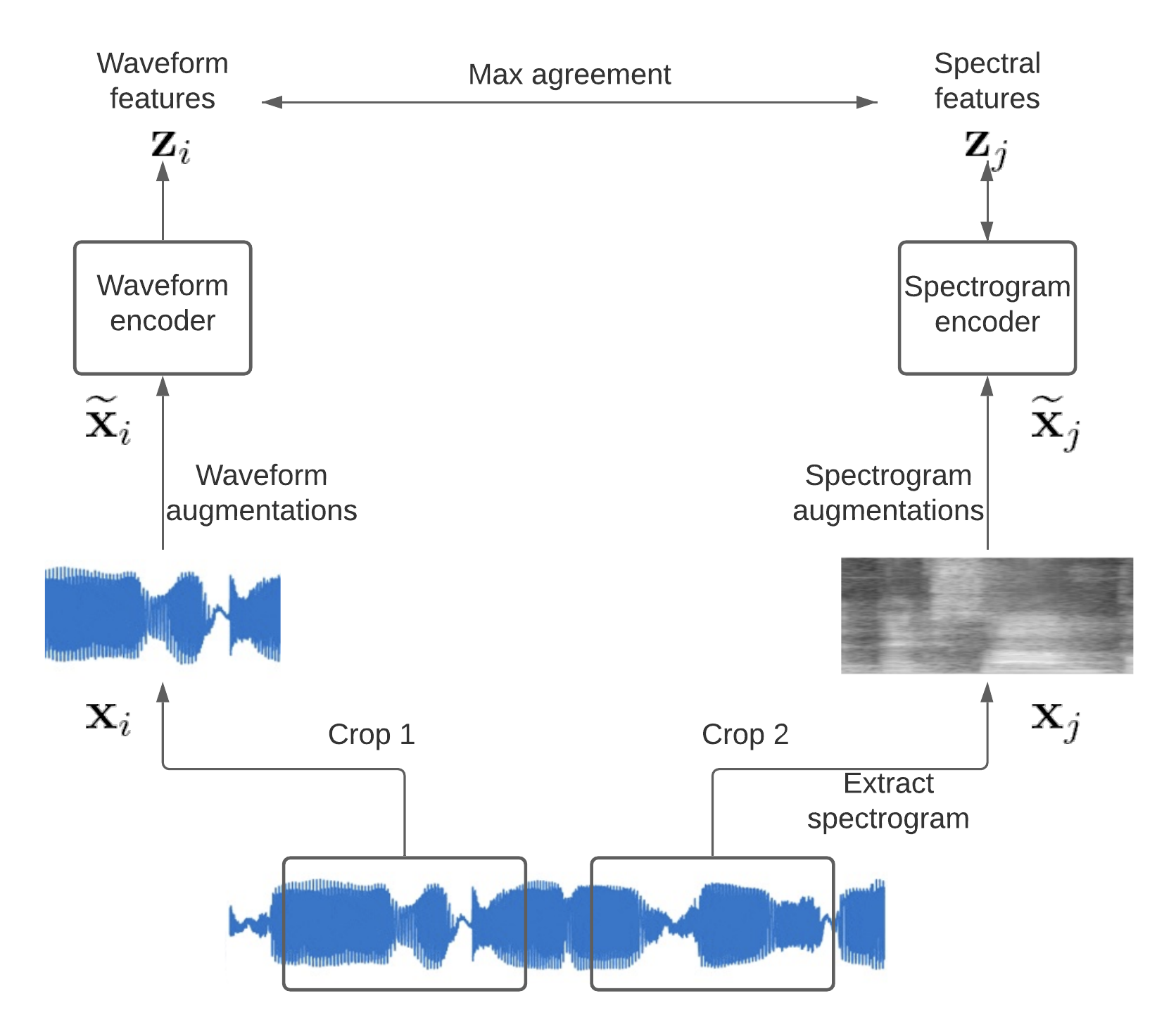}
  \caption{Illustration of the multi-format contrastive audio learning framework.}
  \label{fig:framework}
\end{figure}

The multi-format contrastive audio learning framework is depicted in Figure~\ref{fig:framework}. The input audio in the waveform format is first cropped into two shorter clips $\mathbf{x}_i$ and $\mathbf{x}_j$. One of them can be further transformed into the spectral representation. Waveform or spectrogram augmentations are then applied accordingly, creating the positive pair $\left(\mathbf{\widetilde{x}}_i, \mathbf{\widetilde{x}}_j\right)$. There are different ways to form the distractors or negatives including sampling from the same or other data samples. In this work we choose the negative pairs composing views from two different samples the same way as in SimCLR \cite{chen2020simple}. Then it learns by maximizing the similarity between the encoded representations of the positive pair $\left(\mathbf{z}_i, \mathbf{z}_j\right)$. The loss function is defined as

\begin{equation}\label{eq1}
  L_{i,j} = -\textup{log} \frac{\exp{\left( \textup{sim} \left( \mathbf{z}_i, \mathbf{z}_j \right) / \tau \right)} }{\sum_{k \ne i} \exp{\left( \textup{sim} \left( \mathbf{z}_i, \mathbf{z}_k \right) / \tau \right)} }
\end{equation}

where $\tau$ denotes the temperature parameter, and $\textup{sim} \left( \cdot, \cdot \right)$ is the nonlinear cosine similarity measure with the form of $\textup{sim}\left ( \mathbf{u}, \mathbf{v} \right ) = g\left ( \mathbf{u} \right ) \cdot g\left ( \mathbf{v} \right ) / \left \|  g\left ( \mathbf{u} \right ) \right \| \left \|  g\left ( \mathbf{v} \right ) \right \|$, in which the projector $g$ is a multi-layer perceptron (MLP) model with 1 hidden layer and ReLU nonlinearity. It is shared by both branches.
The summation in the denominator over $k$ is computed from $2N - 1$ crops in the batch (excluding $\mathbf{z}_i$).
Both $L_{i,j}$ and $L_{j,i}$ are computed and summed up as the overall loss for the positive pair $\left(\mathbf{\widetilde{x}}_i, \mathbf{\widetilde{x}}_j\right)$. The final loss is computed across all positive pairs in the batch.

\subsection{Architecture}

For the spectral input format (including spectrograms, log-mel, and MFCCs), we adopt the state-of-the-art supervised models from \cite{kong2020panns} by removing the last two linear layers, and directly use the outputs from the global pooling layer. These models include CNN6, CNN10, CNN14, ResNet22, Resnet38, and ResNet54. For convenience, in this paper we use their original names even though there are two less layers.

When raw audio is presented, we employ networks previously used as the encoders in various unsupervised speech models \cite{schneider2019wav2vec, kawakami2020learning, baevski2020wav2vec}. The building block is a 1D convolutional layer followed by Group Normalization and ReLU activation. The first layer has a kernel size of 10 and stride 5, followed by 5 to 9 layers of kernel size of 4 and stride 2. These models shrink the temporal dimension by 160 to 2560 times, so that we refer them as Conv160, Conv320, ..., Conv2560 in this work. The number of filters in each layer is 512. Global average pooling is applied on the time dimension at the end. Besides, we consider the Res1dNet-31 and Res1dNet-51 model from \cite{kong2020panns}. We remove the last two linear layers and find it is important to use Group Normalizations for these models.

The CNN6, CNN10 and Conv160 to Conv2560 model output a feature space of 512 dimensions, and the rest, namely, CNN14, ResNet22, ResNet38, ResNet54, Res1dNet-31, and Res1dNet-51, result in 2048 dimensions. The latent sizes need to be matched for the loss function when two different models are used. Note that the features from the encoders are pooled from the time and/or frequency dimension. In the ablation studies, if only one audio format is used, we let the two branches share the same model. We have also experimented training two models and concatenate the output features but found it yields similar results. If two different formats are presented, we use two networks and concatenate the features for the downstream tasks. In the final results, we also report the performance of each network.

\subsection{Audio augmentations}

It is observed previously that augmentations are important to the contrastive learning framework \cite{bachman2019learning, henaff2020data, chen2020simple}. We consider different types of augmentations for raw audio and spectrograms \cite{jansen2018unsupervised, park2019specaugment, kharitonov2020data}. As the pitch shift requires additional STFTs on the fly, in this work we consider the following less expensive augmentations:

\paragraph{Audio mixing}

Small additive noise of any sort will not alter the original categories of the audio. Given two audio clips $\mathbf{x_1}$ and $\mathbf{x_2}$, the mixed-up version is

\begin{equation}\label{eq2}
  \mathbf{\hat{x}_1} = \alpha \mathbf{x}_1 + (1 - \alpha) \mathbf{x}_2
\end{equation}

where $\mathbf{\hat{x}_1}$ inheritances labels from $\mathbf{x_1}$. In this work, $\alpha$ is samples from $\beta(5, 2)$ distribution. This simulates various realistic noise conditions.

\paragraph{Time masking}

$t$ consecutive time steps $[t_0, t_0 + t)$ of the audio can be dropped out and it should not change the event classes, where $t_0$ is randomly sampled. This can be applied both to raw audio and spectrograms.

\paragraph{Frequency masking}

A small amount of $f$ frequency components $[f_0, f_0 + f)$ on the spectrogram can be masked out without losing semantic information.

\paragraph{Frequency shift}

One can apply the truncated shift in frequency to the spectrograms by an integer number sampled from $[-F, F]$, where $F$ is the maximum shift size. Missing values after the shift are set to zero energy. Intuitively, this is a less expensive alternative of changing the pitch of the audio.

\section{Experiments}
\label{experiments}

We use the audio segments sampled at 16k Hz from AudioSet \cite{gemmeke2017audio} for both training and evaluation. We split the original training set into training and validation subset by $95\%$ and $5\%$, respectively. We evaluate the representations in the downstream task of training shallow fully connected audio classifiers following the same setup as in \cite{jansen2018unsupervised, jansen2019coincidence}, where a 1-hidden-layer MLP with 512 units is used and the parameters in the pretrained network are fixed. In this section we detail some of key factors that affect the model performance based on the development set. Then we show how the proposed method compares to the state of the art on the test set. 


Unless noted, the models are trained up to 400k steps with a batch size of 1024. Adam optimizer is used, starting from an initial learning rate of $10^{-4}$, and follows a cosine learning rate decay down to $10^{-6}$. We randomly crop two windows of 3 seconds from each data sample during training. On evaluation we equally split the data into overlapped subclips with the stride of half of the crop size, and average the logits from the subclips to obtain the overall score of the clip. The loss has a temperature of 0.1. When the spectral representations are used, they are extracted by a window size of 20 ms and stride of 10 ms. The spectrogram and log-mel features have 80 dimensions. For MFCCs we follow the convention and take 13 features \cite{jurafsky2008speech}. The default \emph{small} model uses CNN10 for spectrograms and Conv320 for raw input, and the \emph{base} model employs CNN14 and Res1dNet-31.

\begin{figure}[t]
  \centering
  \includegraphics[width=0.28\linewidth]{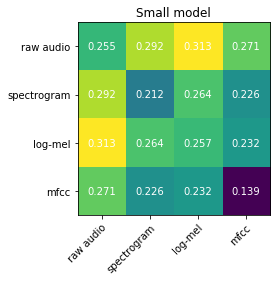}
  \includegraphics[width=0.28\linewidth]{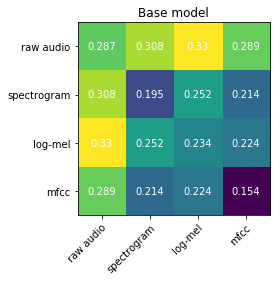}
  \caption{Validation mAP of the small (left) and base (right) models with different combinations of audio formats as input.}
  \label{fig:audio_formats}
\end{figure}

\begin{figure}[t]
  \centering
  \includegraphics[width=0.41\linewidth]{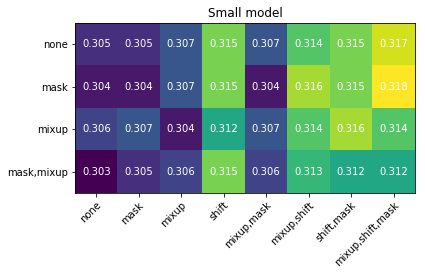}
  \includegraphics[width=0.41\linewidth]{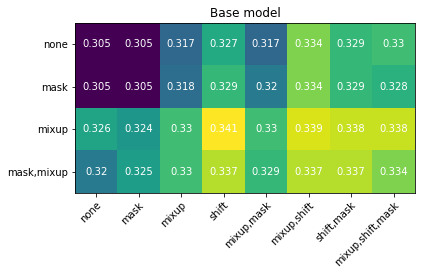}
  \caption{Validation mAP of the raw-audio-vs-log-mel models with different combinations of raw audio (along rows) and spectrogram (along columns) augmentations.}
  \label{fig:augmentations}
\end{figure}

\subsection{Audio formats}

We compare the performances of learning with different combinations of audio formats in Figure~\ref{fig:audio_formats}. It is noticed that both the small and base model benefit from using two kinds of input formats. Meanwhile, maximizing agreement between raw waveforms and the frequency representations outperforms combinations with two spectrograms by a large margin. In particular, the combination of raw audio and log mel spectrograms really stands out: on the base model, it improves relatively upon the raw-audio-only and log-mel-only counterpart by $15\%$ and $41\%$, respectively. The MFCC-based models have the lowest scores, possibly because they are low in feature dimensions. We think the reason why this works well is because taking another format of audio can be viewed as an aggressive way of transforming or augmenting the data to create semantically related but vastly different views, such that the contrastive learning framework can not leverage the trivial cues to solve the proxy task without learning meaningful representations.

\subsection{Creation of the views}

Table~\ref{tab:crop} shows that the model is better trained when taking two randomly cropped clips of 3 to 5 seconds to create the views. Taking the full length (10 seconds) results in the worst performance. This is possibly due to the multi-instance nature of AudioSet. Because our models output features averaged on the time dimension, and some class may only last a very short duration within the clip, taking a long temporal scale may completely bury the short-lasting classes, resulting in a lower validation score. In addition, it also shows that the maximum frequency shift is optimal around half of the frequency dimension size.

In Figure~\ref{fig:augmentations} it is seen that the frequency shift has the biggest impact on both the small and base model. The base model benefits more from the audio mixing, possibly because the small model does not have enough capacity to account for it. In our experiments, masking on either time or frequency does not improve the downstream performance.

\begin{table}[t]
  \caption{Effects of different crop sizes, maximum frequency shifts, temperature parameters, and projection latent sizes on the base model when trained with both raw audios and log mel spectrograms.}
  \label{tab:crop}
  \centering
  \begin{tabular}{ccccccccc}
    \toprule
    Crop size (s) & 1 & 2 & 3 & 4 & 5 & 6 & 8 & 10 \\
    \midrule
    Val mAP & 0.310 & 0.336 & 0.340 & 0.344 & 0.341 & 0.328 & 0.305 & 0.262    \\
    \bottomrule
  \end{tabular}
  \begin{tabular}{cccccccccc}
    \toprule
    Max freq shift & 0 & 2 & 4  & 10 & 20 & 40 & 60 & 80 \\
    \midrule
    Val mAP & 0.331 & 0.329 & 0.333 & 0.340 & 0.340 & 0.342  & 0.338 & 0.326    \\
    \bottomrule
  \end{tabular}
  \begin{tabular}{ccccccc}
    \toprule
    Temperature & 0.05 & 0.1 & 0.25 & 0.5 & 0.75 & 1 \\
    \midrule
    Val mAP & 0.330 & 0.340 & 0.326 & 0.312 & 0.297 & 0.285    \\
    \bottomrule
  \end{tabular}
  \begin{tabular}{cccccc}
    \toprule
    Latent size & 128 & 256 & 512 & 1024 & 2048 \\
    \midrule
    Val mAP & 0.331 & 0.338 & 0.340 & 0.342 & 0.345    \\
    \bottomrule
  \end{tabular}
\end{table}



\subsection{Model architectures}

We run ablations on the choice of model architectures and the results are shown in Table~\ref{tab:model}. For smaller models, we observe gains by increasing the model capacity. However, the same does not hold for large models - the performance does not further improve when the model size goes beyond Res1dNet-31 or CNN14. The same behavior has also been documented in the supervised setting \cite{kong2020panns}. We leave how to scale it further for future work.

\begin{table}[t]
  \caption{Impacts of network architectures when trained with both raw waveforms and log mel spectrograms.}
  \label{tab:model}
  \centering
  \begin{tabular}{cccccc}
    \toprule
    \textbf{Small model} & Conv160     & Conv320 & Conv640 & Conv1280 & Conv2560 \\
    \midrule
    CNN6 & 0.300  & 0.305 & 0.313 & 0.314 & 0.314    \\
    CNN10     & 0.303 & 0.313 & 0.320 & 0.320 & 0.322      \\
    \bottomrule
  \end{tabular}
  \begin{tabular}{ccc}
    \toprule
    \textbf{Base model}    & Res1dNet-31     & Res1dNet-51 \\
    \midrule
    CNN14 & 0.340  & 0.340     \\
    ResNet-22     & 0.335 & 0.336      \\
    ResNet-38     & 0.332       & 0.333  \\
    ResNet-54     & 0.335       & 0.340  \\
    \bottomrule
  \end{tabular}
\end{table}

\subsection{More ablations}

\begin{figure}[ht]
  \centering
  \includegraphics[width=0.53\linewidth]{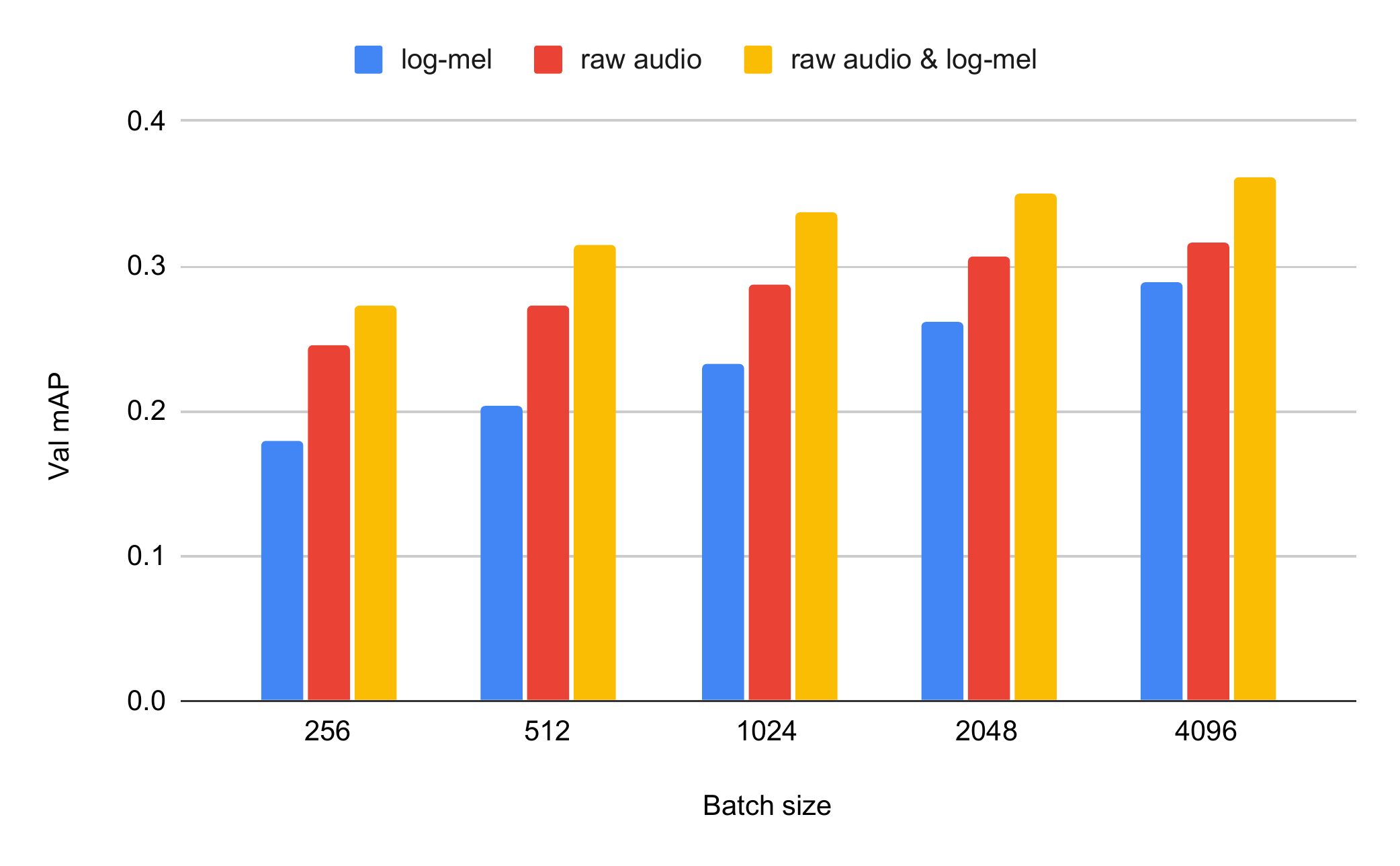}
  \caption{Validation mean average precision of different batch sizes on the base model.}
  \label{fig:batch_size}
\end{figure}

We observe trends similar to the SimCLR image model \cite{chen2020simple} on the choices of the temperature parameter (Table~\ref{tab:crop}), projection MLP latent size (Table~\ref{tab:crop}), batch size (Figure~\ref{fig:batch_size}). In particular, we also find that it is better to train the contrastive framework with a very large batch size, possibly because it needs a large pool of negatives for the softmax loss in Equation~\ref{eq1} to pick out the hard ones.




\subsection{Comparison to the state of the art}

Based on the findings from the ablations above, we train our final model with CNN14 and Res1dNet-31 using a large batch size of 32768. Frequency shift, audio mixing, time and frequency masking are applied to the log-mel branch (CNN14), and audio mixing is used for the raw waveform (Res1dNet-31). It takes 3-second crops and runs up to 700k steps and save the model with the best validation score. The latent size in the projection head is 1024.

We report the test scores on AudioSet in Table~\ref{tab:benckmark}. When only one audio format is used in the contrastive learning framework, it is observed that the results are already better than the previous best score trained with multiple modalities. Specifically, our log-mel-only model has a test mAP of 0.329 and the waveform-only one scores 0.336, outperforming the multimodal versatile network at 0.309, which is trained with audio, video, and texts \cite{alayrac2020self}. This is counter-intuitive because previous works have shown multi-modal learning is better than the single-modal ones. We think this is at least partially because using only the audio modality allows one to train models using the contrastive loss with a very large batch size and more training steps given the same computation budget.

When two audio formats are presented and both the log-mel and waveform network are trained simultaneously, it is seen that each individual network performs better than when trained with only one format. Note that the log-mel network, performing at 0.368 mAP, employs the CNN14 architecture. The same network trained under the supervised setting is reported with a mAP of 0.375 in \cite{kong2020panns} (without class balancing). If we use the  concatenated features from the two networks, the performance further increases to 0.376. The current supervised state of the art of 0.439 mAP is achieved by the Wavegram model trained with both log-mel spectrograms and waveforms \cite{kong2020panns}. However, it is noted that class balancing is crucial to this model, which requires the access to class labels. The self-supervised framework used in this work does not require any label and has no assumption about the class distribution. This ensures that this self-supervised method can scale well with the large amount of unlabelled data. We leave this for future work.

We also present in Table~\ref{tab:esc50_benckmark} the results on generalization to a smaller downstream dataset ESC-50 \cite{piczak2015esc}. It is widely used for evaluating audio representations trained by various cross-modal frameworks. Our multi-format model achieves a new SOTA accuracy of $90.5\%$ without requiring any additional modality or dataset for pre-training. Moreover, without feature concatenation, the raw audio network of this model alone has an accuracy of $89.3\%$, while the log mel network achieves $89.7\%$, which are higher than the corresponding results of $84.9\%$ and $86.3\%$ using single-format training, respectively.

\begin{table}[t]
  \caption{Test performance of shallow model classification on AudioSet with fixed representations.}
  \label{tab:benckmark}
  \centering
  \begin{tabular}{ c c c c }
    \toprule
    \textbf{Model} & \textbf{Train inputs} & \textbf{Eval inputs} & \textbf{Test mAP} \\
    \midrule
    Triplet \cite{jansen2018unsupervised} & log-mel  & log-mel        & $0.244$~~~\\
    $L^3$ \cite{arandjelovic2017look} & log-mel $+$ video & log-mel & $0.249$~~~\\
    CPC \cite{wang2020contrastive}                  & waveform   & waveform       & $0.277$~~~\\
    $C^3$ \cite{jansen2019coincidence} & log-mel $+$ video & log-mel  & $0.285$~~~\\
    MMV \cite{alayrac2020self} & log-mel $+$ video $+$ text & log-mel & $0.309$~~~\\
    \midrule
    Ours                  & log-mel   & log-mel       & $0.329$~~~\\
    Ours                  & waveform  & waveform        & $0.336$~~~\\
    Ours                  & waveform $+$ log-mel  & log-mel        & $0.368$~~~\\
    Ours                  & waveform $+$ log-mel  & waveform        & $0.355$~~~\\
    Ours                  & waveform $+$ log-mel  & waveform $+$ log-mel        & $\mathbf{0.376}$~~~\\
    \midrule
    \spv{Supervised \cite{kong2020panns}} & \spv{waveform $+$ log-mel} &  \spv{waveform $+$ log-mel}  & \spv{$0.439$}~~~\\
    \bottomrule
  \end{tabular}
  
\end{table}

\begin{table}[t]
  \caption{Test accuracy of linear classification on ESC-50 with fixed audio representations. Hyperparameters of the classifier are selected with split 1 and the average accuracy over 5 splits is reported.}
  \label{tab:esc50_benckmark}
  \centering
  \begin{tabular}{ c c c c }
    \toprule
    \textbf{Model} & \textbf{Train inputs} & \textbf{Eval inputs} & \textbf{Test accuracy ($\%$)} \\
    \midrule
    $L^3$ \cite{arandjelovic2017look} & log-mel $+$ video & log-mel & $79.3$~~~\\
    AVTS \cite{korbar2018cooperative} & log-mel $+$ video & log-mel  & $82.3$~~~\\
    XDC \cite{alwassel2019self} & log-mel $+$ video & log-mel & $84.8$~~~\\
    GDT \cite{mandela2020datatrans} & log-mel $+$ video & log-mel & $88.5$~~~\\
    MMV \cite{alayrac2020self} & log-mel $+$ video $+$ text & log-mel  & $88.9$~~~\\
    AVID \cite{morgado20avid}  & log-mel $+$ video & log-mel & $89.2$~~~\\
    \midrule
    Ours                  & log-mel     & log-mel     & $86.3$~~~\\
    Ours                  & waveform    & waveform      & $84.9$~~~\\
    Ours                  & waveform $+$ log-mel & log-mel & $89.7$~~~\\
    Ours                  & waveform $+$ log-mel & waveform & $89.3$~~~\\
    Ours                  & waveform $+$ log-mel & waveform $+$ log-mel & $\mathbf{90.5}$~~~\\
    \midrule
    \spv{Supervised \cite{kong2020panns}}  &  \spv{waveform $+$ log-mel} & \spv{log-mel}  & \spv{$90.8$}~~~\\
    \bottomrule
  \end{tabular}
  
\end{table}

\section{Conclusions}

In this work, we study the use of multiple formats of the audio for contrastive learning. We observe a significant advantage when training with both raw waveforms and log-mel spectrograms. Our model improves the state of the art on the AudioSet benchmark relatively by $21.7\%$, bridging the gap between unsupervised and supervised learning. Our work shows that multi-format training is promising to fully unlock the potential of large-scale and audio-only unsupervised learning.

\begin{ack}
The authors would like to thank Yan Wu for fruitful discussions. We also appreciate the feedback from the anonymous reviewers.
\end{ack}




\bibliographystyle{unsrt}
\bibliography{refs}

\end{document}